\documentclass[twocolumn,aps,pra,longbibliography,showpacs,cnofootinbib,
superscriptaddress,10pt]{revtex4-1}
\usepackage{eurosym}
\usepackage{amsfonts}
\usepackage{amsmath}
\usepackage{amssymb}
\usepackage{graphicx}
\usepackage[latin1]{inputenc}
\usepackage{graphicx,dcolumn,bm}
\usepackage{multirow}
\usepackage{array}
\usepackage{float}
\usepackage{subfigure}
\usepackage{color}
\usepackage[colorlinks,hyperindex]{hyperref}
\usepackage{amsfonts}
\usepackage{amssymb}
\usepackage{amsmath}
\usepackage{latexsym}

\begin{document}

\title{Steady many-body entanglements in dissipative systems}
\author{G. D. de Moraes Neto}
\affiliation {Instituto  de F\'{i}sica, Universidade Federal de 
Uberl\^{a}ndia, Uberl\^{a}ndia, MG 38400-902, Brazil}

\author{V. F. Teizen}
\affiliation{Instituto de F\'isica de S\~ao Carlos, Universidade de S\~ao 
Paulo, S\~ao Carlos, S\~ao Paulo 13560-970, Brazil}
\author{V. Montenegro}
\affiliation{Facultad de F\'isica, Pontificia Universidad Catolica de Chile, 
Casilla 306, Santiago, Chile}

\author{E. Vernek}
\affiliation {Instituto  de F\'{i}sica, Universidade Federal de 
Uberl\^{a}ndia, Uberl\^{a}ndia, MG 38400-902, Brazil}
\begin{abstract}
We propose a dissipative method for the  preparation of many-body steady entangled states in spin and fermionic chains. The scheme is accomplished by means of an engineered set of Lindbladians acting over the eigenmodes of the system, whose spectrum is assumed to be resolvable. We apply this idea to prepare a particular entangled state of a spin chain described by the $XY$ model, emphasizing  its generality and experimental feasibility. Our results show that our proposal  is capable of achieving high fidelities and purities for a given target state even when  dephasing and thermal dissipative processes are taken into account. Moreover, the method exhibits a remarkable robustness against fluctuations in the model parameters. 
\end{abstract}
\maketitle

\affiliation{Instituto de F\'{i}sica, 
Universidade Federal de Uberl\^{a}ndia, Uberl\^{a}ndia, Minas Gerais 38400-902, 
Brazil} 

\affiliation{Instituto de F\'{i}sica, Universidade Federal de 
Uberl\^{a}ndia, Uberl\^{a}ndia, Minas Gerais 38400-902, Brazil}

\affiliation{Instituto de F\'{i}sica de S\~ao Carlos,  
Universidade de S\~ao Paulo, 13560-970 S\~ao Carlos, SP, Brazil}

\affiliation{Department of Physics and Astronomy,  
University College London, Gower Street, London WC1E 6BT, United 
Kingdom}%
\affiliation{Instituto de F\'{\i}sica, Pontificia Universidad 
Cat\'{o}lica de Chile,
Casilla 306, Santiago, Chile}

\section{Introduction}

Deterministic state  preparations involving engineered Hamiltonians have promoted a great progress in quantum information in the recent years \cite{detprepper}. Nevertheless, such an endeavour is unavoidably affected by several sources of quantum noise or decoherence, for instance, energy losses from the system of interest to the environment\cite{Livro,revorsz}. The development of strategies to prepare nonclassical states \cite{PNS} and, particularly, to circumvent their decoherence  has long been a challenge in quantum information studies. Despite this, important efforts have been proposed to overcome this obstacle , e.g., via decoherence-free subspaces \cite{DFS, destorsz}, dynamical decoupling \cite{DD}, and reservoir engineering \cite{PCZ,RE}. From the conceptual viewpoint, the need for these states stems from their use in the study of fundamental quantum processes, such as decoherences \cite{Decoherence} and quantum-to-classical transitions\cite{QC}. For practical purposes, on the other hand, the advent of the quantum computation and quantum communication fields --- which depends strongly on the production of long-lived quantum states and quantum correlations \cite{Livro, fiber} --- has certainly demanded extra efforts from researchers to develop efficient techniques for preparing and protecting nonclassical states from quantum noise \cite{Orsz}.

In this context, the reservoir  engineering technique proposed in Poyatos et. al. \cite{PCZ} and experimentally demonstrated in a trapped ion system \cite{IT} signals an important step towards the implementation of quantum information resources \cite{Livro}. One of the most important aspects of dissipative protocols is their independence on initial states, i.e., starting from an arbitrary initial state, the non-unitary time  evolution of the system renders a final steady state that asymptotically approaches a predefined 
target state.
% 
% can be drive evolves  and great experimental flexibility.
% its % unnecessary control of an unitary dynamics and target state fidelity 
% increase that asymptotically tends to unity, 
% 
Such scheme relies on the construction of a Liouvillian ($\mathcal{L}$) for which the steady state  $(\rho _{S})$ follows the  condition $\mathcal{L}\rho _{S}=0$. If the Liouvillian is engineered in a way that $\rho_S$ is the target state, $\rho_{t} = | \psi_t \rangle \langle \psi_t |$, the dissipative protocol is successful.
% 
% Moreover, 
% the protection of a 
% particular state demands the (not-always-easy) engineering of a specific 
% coupling through which the system of interest is forced to interact 
% with other auxiliary quantum systems. Furthermore,
Furthermore, the reservoir engineering technique can fulfill other purposes such as dissipative preparation of many-body quantum states \cite{Many} and universal dissipative quantum computation \cite{UniDissQC}. Interestingly, this technique is also important for extending the concept of analog quantum simulation to the domain of open systems \cite{ExpDissEng}, allowing for the study of quantum phase trasitions.

The possibility of preparing maximally  entangled states via engineered dissipative processes have been shown both theoretically \cite{re} and experimentally \cite{ExpDissEng}. However, most of the proposed schemes concentrate on the preparation of atomic maximally entangled states of two \cite{2atom} and three qubits ($W$-states) \cite{3atom} and cluster entangled states \cite{Natom}. Extensions to continuous variable cases have been presented to the study of entanglements of two and three oscillators coupled to a common reservoir\cite{2harmonic}. More recently, a proposal for the preparation of steady entanglements in bosonic dissipative networks has also been reported~\cite{Neto2014}. 
Of particular interest for quantum computation and quantum information are the so-called quantum many-body states~\cite{manybody}. Because of the intricate nature of the Hamiltonians describing many-body systems, preparation of many-body quantum states are generally challenging. Recent proposals for preparing such states are based, for instance, on a spin system coupled to a damped harmonic oscillator~\cite{Plenio} or a series of optimized coherent pump pulses followed by feedback operations~\cite{MORIGIWINELAND}. 

In this work we propose a simple yet efficient dissipative  protocol to generate many-body entangled states. Our approach follows the prescription described in \cite{BR}, where the weak-coupling regime between system-reservoir is assumed. The scheme is built up by tunable quantum two-level systems (TLS) with a switchable coupling to a spin system that has one nondegenerate eigenstate as the target state. A suitable initial pump of the TLS to the excited state provides an engineered Liouvillian superoperator that drives the system of interest to the desired steady state (the target state). In order to counterbalance inevitable effects of the natural (nonengineered) environment and the need of polarized initial spin states, we employ other TLSs to cool down the remaining (undesired) eigenstates. Moreover, similarly to the proposals found in Refs.~\cite{Plenio,MORIGIWINELAND}, we need the spectral resolution in the  vicinity of the target state (to avoid spurious population transfer).
% WE NEED TO DISCUSS THESE BEFORE INTRODUCING OURS} 
% However, our proposal is altogether different 
% from \cite{Plenio} where the ideal spin system is coupled to a damped 
% harmonic oscillator and from \cite{MORIGIWINELAND} that relies on a series 
% of optimized coherent pulses followed by feedback operations.
% Although our approach sounds somewhat complementary to the previous studies, 
As a concrete example,  we apply our protocol to prepare a many-body state of a spin chain model\cite{Kay}. Since this model can also be understood as a quadratic fermionic model, the results may also be interesting for fermionic atoms in optical lattices\cite{fermions}.
Experimentally, \textit{``TLS reservoirs"} can be implemented under current technology in quantum cavity electrodynamics~\cite{BR}, in trapped ions \cite{Rafael} and QED superconducting circuits~\cite{Nori}, where a beam of atoms simulating the reservoir can be achieved by a pulsed classical field. In the former, the classical field is used to couple the vibrational field intermittently with the internal ionic states, while the latter, it is used to bring a cooper-pair box into resonance with the mode of a superconducting strip. Moreover, spin chains similar to the ones treated here have already been experimentally implemented in optical lattices\cite{fermions}, trapped ions \cite{ions}, and circuit QED \cite{circuit}. 
Our results reveal that our protocol not only works in the presence of dephasing (which is specially critical for many-body quantum states), thermal effects, and parameter fluctuations, but also has the potential to be scalable. Remarkably, our results show to be very robust against fluctuations in the model parameters as well, which is of paramount importance for experimental realizations. 

\onecolumngrid

\begin{figure*}[h]
\includegraphics[width=0.95\textwidth]{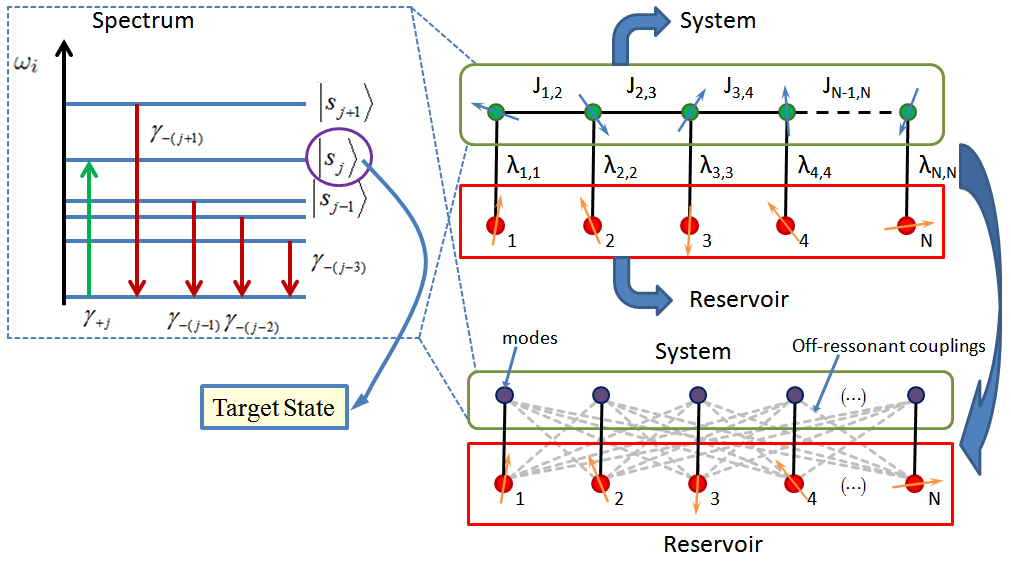}  
\caption{Schematic  representation of the system composed of a spin chain (our system of interest) described by an XY Hamiltonian. The reservoir consists of a set of two level systems that mediates the dissipation and pumping of the spin chain normal modes. By controlling the system-reservoir couplings one can, in principle, find a Hamiltonian whose spectrum is resolved enough thus through its pump we can access the target state ($| s_j \rangle$) individually. $| s_i \rangle$ is the $i$-th eigenstate of the Hamiltonian of the system of interest with $\omega_i$ its corresponding eigenenergy.  $\gamma_{+i}$ $(\gamma_{-i})$ is the pump (cooling) rates of the $i$-th level of the Hamiltonian, $J_{i,i+1} = J$ is the system coupling between spins while $\lambda_{i,k}$ is the coupling between the $i$-th system spin and the $k$-th reservoir spin. In the scheme above $i=k$; the emerging off resonant couplings can be eliminated by performing the rotating wave approximation (RWA).}
\label{fig:fig1}
\end{figure*}

\twocolumngrid

\section{Model}

Our state protection protocol relies on a resonant coupling between a fermionic eigenstate (mode of a spin chain with $N$ elements) and a set of TLSs, being this one, the required engineered reservoir. The setup is schematically depicted in Fig.~\ref{fig:fig1}. The dissipation in those modes is engineered in such a way that the steady state (target state) is an eigenstate of the spin chain Hamiltonian. Thus, the target state is selectively pumped to the relevant spectral gap relative to the ground state of the Hamiltonian - with remaining couplings strongly off-resonant. The steady state of the system of interest is driven by a sum of $N$ engineered Lindbladians, one of which drives the system to the desired target state, while the other $N-1$ Lindbladians are necessary to counterbalance the inevitable effects of amplitude and phase damping, both emmerging naturally from the (nonengineered) environment. We demonstrate that once the system Hamiltonian has been diagonalized (either  analytically or numerically) and both provided that the spectrum is resolvable and the fermionic modes can be directly coupled to the spin reservoir,  our method can be used to address various entangled states. Let us start by defining  the total Hamiltonian $H=H_{c}+H_{r}+H_{I}$, where $H_c$ and $H_r$ correspond to the Hamiltonians of the system and the reservoirs, respectively. Specifically, $H_c$ corresponds to  a spin chain Hamiltonian and can be expressed by
\begin{equation}
H_{c}=\sum_{i,j=1}^N O_{i}^{\dagger }H_{ij}O_{j},
\end{equation}
with $H_{ij} = \omega_{i} \delta_{ij} + \zeta_{ij} (1-\delta_{ij})$, $O_{i}^{\dagger}$ ($O_{i}$) being the creation (annihilation) fermionic operators, $\omega_i$ denoting the $i$-th fermion frequency, and $\zeta_{ij}$ representing the coupling strength between the pair of fermions, and $\delta_{ij}$ the Kronecker delta. The reservoir is composed by a set of TLSs with frequencies $\Omega_j$  whose Hamiltonian, can be written as
\begin{equation}
H_{r}=\sum_{j=1}^{N}\frac{\Omega _{j}}{2}\sigma _{j}^{z},
\end{equation}
in which $\sigma _{j}^{z}$ is the usual  Pauli operator. Finally, the interaction between the system  and the reservoir is described by
\begin{eqnarray}
H_{I}=\sum_{j=1}^{N}\lambda _{j}(O_{j}^{\dagger }\sigma _{j}^{-}+O_{j}\sigma
_{j}^{+}).
\end{eqnarray}
where $\lambda _{j}$ is the strength of the fermion-reservoir coupling when the RWA approximation has already been taken into account. In general, the expression above would contain several time dependent terms, thus in order to overcome this difficulty, we chose detunings in such a way to avoid (small or null) irrelevant couplings and, in this manner, being able to generate the desired target states. The frequencies of the modes and TLSs are already resonant with a relevant transition of interest. 

 %In general this expression would have various time dependent terms and 
% we'll choose such detunnings in such a way as to provide the coupling 
% necessary to generate the target states we're interested in.

We now diagonalize the Hamiltonian $H_{I}$ using  the transformation $R_{m}=\sum_{n}T_{mn}^{-1}O_{n}$, where the coefficients of the $n$-th column of the orthonormal matrix $T$ ($T^{-1}=T^{t}$) gives the eigenvectors associated with the eigenvalues $\tilde{\omega _{i}}$. This allows us to express the Liouvillian in terms of the operator that will protect the target state after tracing out the reservoir degrees of freedom. To engineer quantum states we use the atomic reservoir technique \cite{BR,FockEPL}, in which, traditionally, an atom beam passing through a cavity mode (one atom at a time) is used to generate an artificial Liouvillian. Within such technique,  the weak coupling regime for the interaction parameter of the Hamiltonian is assumed, i. e., $\lambda_{i}\tau <<1$, where $\tau $ is the mean interaction time, yielding the Markovian Liouvillian,
\begin{eqnarray}
\mathcal{L}_{\rm eng}\rho &=&\frac{\Gamma_{-}}{2}\left( 2O\rho O^{\dagger 
}-\rho O^{\dagger }O-O^{\dagger}O\rho \right) \nonumber \\
&+&\frac{\Gamma _{+}}{2}\left( 2O^{\dagger } \rho O-\rho OO^{\dagger 
}-OO^{\dagger }\rho \right) \nonumber.
\end{eqnarray}
In the above, the effective rate $\Gamma _{+}$ ($\Gamma _{-}$) accounts for the pumping (cooling) of the target state $ \rho _{_{T}}$. They are given by \cite{BR} $\Gamma _{\pm }=r_{\pm }(\lambda _{i}\tau )^{2}$, in which $r_{+}(r_{-})$ is the rate of the switching on (off) of the interaction that drives  the TLS to the excited (ground) state. The total Liouvillian also includes dissipative terms associated with thermal and phase losses, which are taken into account in our model as well. In the next section we illustrate this protocol and clarify the principles mentioned above in the $XY$ spin chain Hamiltonian.

For concreteness, let us consider a finite spin chain composed of $N$ spins 
coupled by the isotropic $XY$ Hamiltonian with each spin coupled to a TLS. The 
total Hamiltonian $H=H_{c}+H_{r}+H_{I}$ of this system is given by ($\hbar 
=1$)
\begin{eqnarray}
H_{c} &=&J\sum_{j=1}^{N-1}(S_{j}^{+}S_{j+1}^{-}+S_{j}^{-}S_{j+1}^{+}), \\
H_{r} &=&\sum_{j=1}^{N}\frac{\Omega _{j}}{2}\sigma _{j}^{z}, \\
H_{I} &=&\sum_{j=1}^{N}\lambda _{j}(S_{j}^{+}\sigma
_{j}^{-}+S_{j}^{-}\sigma_{j}^{+}),
\end{eqnarray}%
where $S_{j}^{l}(\sigma _{j}^{l})$  (with $l=+,-$)  denotes the Pauli raising (lowering) operator of the $j-th$ spin (TLS), $J$ is the strength of the nearest neighbor spin-spin coupling within the system of interest, $\lambda_j $ is the strength of the chain-reservoir coupling between the $j$-th spin of the chain and the $j$-th reservoir TLS, $\Omega _{j}$ is the frequency of the $j$-th TLS. Performing a Jordan-Wigner transformation, it is possible to fermionize (and diagonalize) the Hamiltonian of the system of interest, leading to $\bar{H}_{\rm chain}=\sum_{k=1}^{N} \omega _{k}f_{k}^{\dagger }f_{k},$ where $\omega_{k}=2J\cos\left( \frac{k \pi}{N+1}\right) $ \cite{diago}, with $k=1,\cdots ,N$ and $f_{k}^{\dagger},f_{k}$ are, respectively, the creation and annihilation fermionic operators and $\bar{H}_{\rm chain}$ is the diagonalized chain Hamiltonian. With this diagonalized chain term it is possible to define an interaction picture by performing the unitary transformation $U=\exp \left\{ -i\left[\sum_{k=1}^{N}\omega _{k}f_{k}^{\dagger} f_{k}+\sum_{j=1}^{N} \frac{\Omega_{j}}{2}\sigma_{j}^{z}\right] t\right\} $, then $H_{I}$ takes the form,
\begin{eqnarray}
\tilde{H}_{I} =\sum_{j=1}^{N}\sum_{k=1}^{N}\lambda
_{jk} \left\{f_{k}^{\dagger}\sigma _{j}^{-}\exp\left[i(\omega _{k}-\Omega 
_{j})t\right] \nonumber \right. \\
\left.  +f_{k} \sigma 
_{j}^{+}\exp\left[-i(\omega _{k}-\Omega _{j})t\right]\right\},
\end{eqnarray}%
with $\lambda _{jk}= \sqrt{\frac{2}{N+1}}\sin\left(\frac{jk\pi }{N+1}\right) $. We now tune each TLS resonantly with one  eigenmode $\omega _{k}=\Omega_{j}$ (with $k$ and $j$ chosen in order that irrelevant couplings are avoided) with a detuning $\omega_{k\pm 1}-\Omega _{j}\approx J/N$. Assuming  $J/N\gg \lambda_{jk\text{\ }}$  we can write the effective Hamiltonian (within the RWA) as
\begin{eqnarray}
\tilde{H}_{I}=\sum_{i=1}^{N}\lambda _{i}(f_{i}^{\dagger }\sigma
_{i}^{-}+f_{i} \sigma _{i}^{+}).  \label{RWA}
\end{eqnarray}%
The validity of this effective RWA and of the full Hamiltonian has been analyzed in detail in the Refs.~\cite{Neto2016}~and~\cite{Roversi}.

Following Refs.~\cite{BR} and \cite{FockEPL},  we assume a weak-coupling regime for the interaction parameter, i.e., $\lambda_{i}\tau\ll 1$ (with $\tau $ being the time in which the TLS interacts with the spin system). When the TLSs are prepared in the ground $ |g\rangle $ or in the excited $|e\rangle $ states, we obtain the engineered Liouvillian
\begin{eqnarray}
\mathcal{L}_{\rm eng}\rho &=&\sum_{i=1}^{N}\frac{\gamma _{i^{-}}}{2}\left(
2f_{i}\rho f_{i}^{\dagger }-\rho f_{i}^{\dagger }f_{i}-f_{i}^{\dagger
}f_{i}\rho \right)  \notag \\
&&+\sum_{i=1}^{N}\frac{\gamma _{i^{+}}}{2}\left( 2f_{i}^{\dagger }\rho
f_{i}-\rho f_{i}f_{i}^{\dagger }-f_{i}f_{i}^{\dagger }\rho \right) ,
\end{eqnarray}
where $\gamma _{i^{-}}=r_{g}(\lambda _{i}\tau )^{2}$, $\gamma_{i^{+}}=r_{e}(\lambda _{i}\tau )^{2}$ and $r_{g(e)}$ the switching on/off rate of the TLS-Spin interaction.The dissipative dynamics of the open system is assumed  to be Markovian and governed by a master equation of Lindblad form
\begin{eqnarray}
\mathcal{L}_{nat}\rho &=&\sum_{i}\frac{\kappa }{2}(1+\overline{n} ) \left( 2S_{i}^{-}\rho S_{i}^{+}-\rho
S_{i}^{+}S_{i}^{-}-S_{i}^{+}S_{i}^{-}\rho \right)  \notag \\
&&+\sum_{i}\frac{\kappa }{2}\overline{n} \left( 2S_{i}^{+}\rho
S_{i}^{-}-\rho S_{i}^{-}S_{i}^{+}-S_{i}^{-}S_{i}^{+}\rho \right)  \notag \\
&&+\sum_{i} \frac{\kappa _{\phi }}{2}\left( 2S_{i}^{z}\rho S_{i}^{z}-\rho
\right) , 
\end{eqnarray}
with decay rates $\kappa $, dephasing $\kappa _{\phi}$  and $\overline{n}$ is the average number of bosons in the thermal bath. We want to drive the system of interest to a target steady state. To that effect, we choose an eigenmode of the system of interest and prepare the associated reservoir in $|e\rangle $ while the other $N-1$ \textit{``TLS reservoirs"} are prepared in the $ |g\rangle $ - and, as it will be shown later in the numerical results section, it might not be necessary to use all of the $N-1$ depumping terms. To validate our protocol we solve numerically the full master equation
\begin{equation}
\frac{d\rho }{dt}=- i \left[ \tilde{H}_{I},\rho \right]  
+\mathcal{L}_{\rm nat}\rho +\mathcal{L}_{\rm eng}\rho, 
\end{equation}
running in QuTIP \cite{QuTIP}, and compute the steady-state density matrix $\rho _{S}$ as $t\rightarrow \infty.$
\section{Numerical Results}
\begin{figure*}[t!]
\centering
\includegraphics[width=1.0\textwidth]{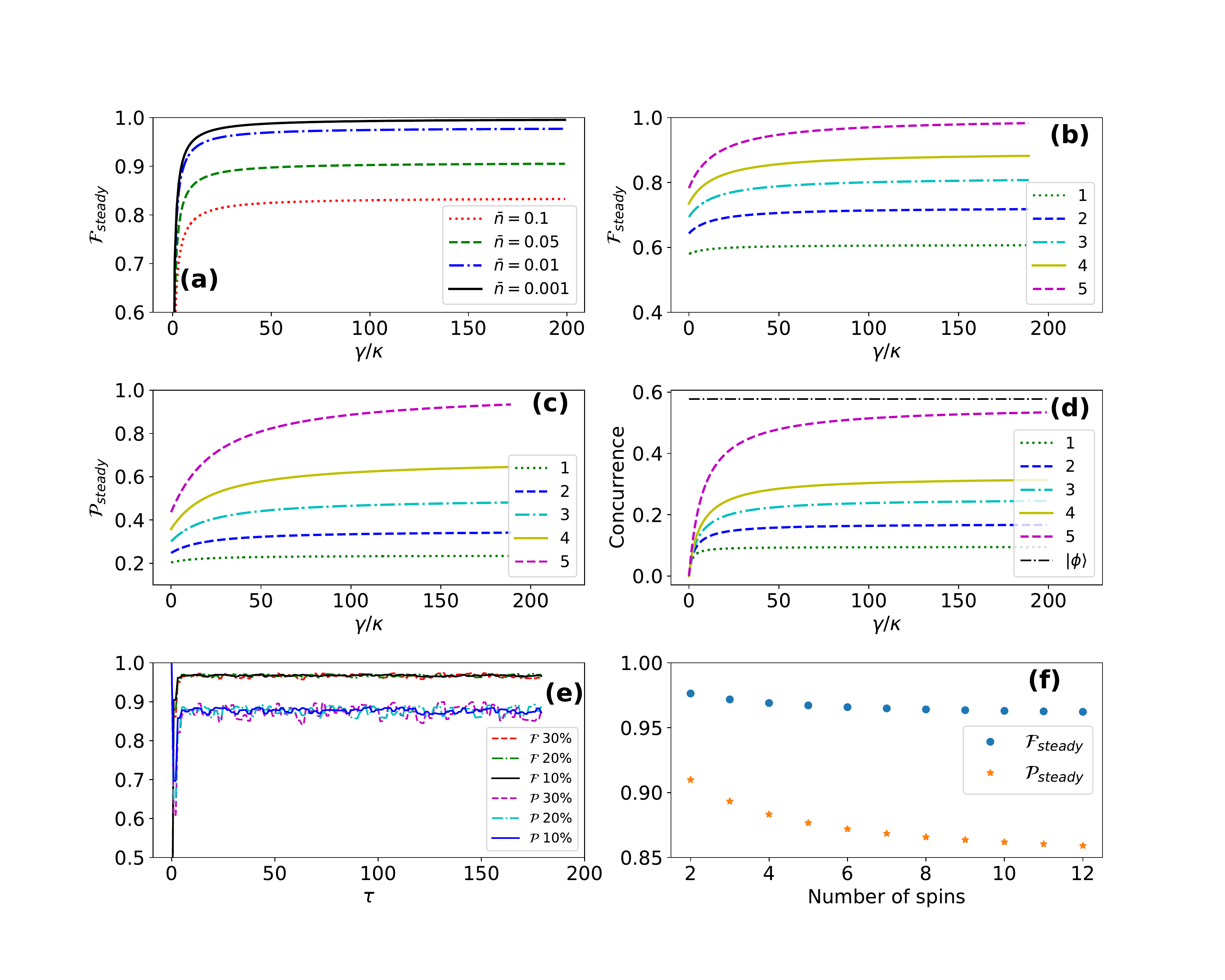}  
\caption{(a) Fidelity $\mathcal{F}_{\rm steady}=\protect\sqrt{Tr ( \left\vert \protect\phi \right\rangle \left\langle \protect\phi \right\vert \protect\rho _{ss} ) }$ as a function of $\protect\gamma /\protect \kappa$ for various temperatures (represented by $\bar n$) for the $N=5$ spin chain, with steady state as given by the Eq.~\eqref{state_phi}. Here we consider only one engineered reservoir connected with the pump Liouvillian at $\protect\gamma_{^{+}}=\protect\gamma $ and  $\protect\gamma_{^{-}}= \protect\kappa_{\protect\phi }=0$. (b)  Same as in (a) but for five  different number of reservoirs (as indicated in the figure). Here we set $\protect\gamma _{^{+}}=\protect\gamma _{^{-}}=\protect\gamma $, $\protect\kappa_{\protect\phi }=\protect\kappa $. The temperature of all the reservoirs is assumed to be equal, $\overline{n} =0.001$. (c)  purity $\mathcal{P}_{steady}=Tr\left[ \rho _{ss}^{2}\right]$ and (d) concurrence between the second and third spins against $\protect\gamma /\protect \kappa$ for the same parameters of panel (b). (e) Time  evolution ($\tau =t/\protect\gamma$) of the fidelity  and purity $\mathcal{P}(t)$  for three  different degrees of randomness $(10,20,30)\%$ of the effective decay rates $\protect\gamma_{^{+}}$ and $\protect\gamma_{^{-}}$. (f) Fidelity and Purity are obtained for the state $\left\vert \phi \right\rangle _{N}$ against number $N$ of spins in the chain with the other parameters kept the same as in the panel (e).}
\label{fig:loctermico3}
\end{figure*}

Let us start by considering the effect of the temperature on the preparation of the steady (entangled) state
\begin{eqnarray}\label{state_phi}
\left\vert \phi \right\rangle &=&\frac{1}{2\sqrt{3}}  \left\vert \mathbf{\uparrow \downarrow \downarrow \downarrow \downarrow } \right\rangle +\frac{1}{2}\left\vert \downarrow \uparrow \downarrow \downarrow \downarrow \right\rangle +\frac{1}{\sqrt{3}}\left\vert \downarrow\downarrow \uparrow \downarrow \downarrow \right\rangle \nonumber \\
&+&\frac{1}{2} \left\vert \downarrow \downarrow \downarrow \uparrow \downarrow \right\rangle +\frac{1}{2\sqrt{3}}\left\vert \downarrow \downarrow \downarrow \downarrow \uparrow \right\rangle 
\end{eqnarray}
in a chain of $N=5,$ when only one engineered  reservoir is present, i.e., only the pump Liouvillian with $\gamma _{^{+}}=\gamma $ and $\gamma _{^{-}}=\kappa_{\phi }=0$. To show that indeed the system approaches the target state, in Fig.~\ref{fig:loctermico3}(a) we show the fidelity  $\mathcal{F}_{\rm steady}=\sqrt{Tr\left\vert \phi \right\rangle \left\langle \phi \right\vert \rho _{ss}}$ for which the steady state is obtained for different $\overline{n} $ as a function of $ \gamma /\kappa .$ We observe that, for typical qubits (atoms, ions, superconductors) on microwave experiments, where $\overline{n} \approx 0.001$ \cite{reviewEXP}, we can achieve a fidelity above $0.9$ with $\gamma /\kappa \approx 50$ and $r_{e}\approx 10^{3},$ as currently done in cavity QED~\cite{Haroche}. To clarify the role played by the $N-1$ cooling Liouvillians, and to show how our protocol improves over previous works, in Fig.~\ref{fig:loctermico3}(b) we plot the steady state fidelity against $\gamma /\kappa $ for different numbers of engineered reservoirs. Here we choose $\gamma _{^{+}}=\gamma _{^{-}}=\gamma $, $\kappa _{\phi }=\kappa $ and $\overline{n} =0.001$. Note that when only one pump is used the fidelity decreases to $0.6$ even for high $\gamma /\kappa $ rates.  This reveals that the  steady state fidelity is strongly sensitive to dephasing. That being said, our scheme shows how to circumvent this unwanted effect by adding other reservoirs acting to cooldown the undesired modes. Therefore, when all the reservoirs are turned on we can obtain $\mathcal{F}_{\rm steady}$ close to unity (see for, instance, the magenta line for 5 reservoirs). It is known that the fidelity alone is not enough to asses the quality of the steady state. Therefore, in Fig.~\ref{fig:loctermico3}(c) and \ref{fig:loctermico3}(d) (with the same parameters of Fig.~\ref{fig:loctermico3}(b)), we show  the purity $\mathcal{P}_{\rm steady}=Tr\left[ \rho _{ss}^{2}\right] $, as a complementary measure since it quantifies the degree of mixing. In Fig.~\ref{fig:loctermico3}(d) we compute the concurrence~\cite{con} between the second and third spins. As pointed out, the $\mathcal{F}_{\rm steady}$ is dubious as a witness of merit (one can have high fidelity for a fairly mixed state, as a comparison between figures (b) and (c) reveals upon inspection), as we see high purity only with all the five reservoirs turned on. In this case, we achieved $\mathcal{P}_{\rm steady}$ above $0.9$ and concurrence close to that of the ideal target state $ \left\vert \phi 
\right\rangle$.

An important feature for experimental realizations is to test its robusteness against variations on the system parameters. To analyze this, in Fig.~\ref{fig:loctermico3}(e) we use the same parameters of Fig.~\ref{fig:loctermico3}(b) with all Liouvillians switched on, setting $\gamma /\kappa =10^{2}$. We show the evolution of $\mathcal{F}(t)=\sqrt{Tr ( \left\vert \phi \right\rangle \left\langle \phi \right\vert \rho (t)\text{ } )}$ and $\mathcal{P}(t)=Tr\left[ \rho (t)^{2}\right] $ versus the scaled time $\tau =t/\gamma $, for three different degrees of randomness $(10,20,30)\%$ on the effective decay rates $ \gamma _{^{+}}$ and $\gamma _{^{-}}$. In this scenario, we are effectively introducing fluctuations in the interaction parameter $\lambda \tau $, as well as in the rates $r_{+}(r_{-})$ the switching on (off) the interaction. As displayed in Fig.~\ref{fig:loctermico3}(e), despite the wide range of parameter fluctuation, our scheme shows to be very robust within the considered random parameter fluctuations. Finally, in Fig.~\ref{fig:loctermico3}(f) we investigate the scalability of our protocol, and once again we plot the $\mathcal{ F_ { \mathrm{steady} }}$ and $\mathcal{P_{\mathrm{steady}}}$ to obtain the target state $\left\vert \phi \right\rangle _{N}=f_{1}^{\dagger }\left\vert 0 \right \rangle _{N}=\sqrt{\frac{2}{N+1}}\sum_{j=1}^{N}\sin \left( \frac{j\pi }{N+1}\right) S_{j}^{+}\left\vert 0\right\rangle _{N}$, where $\left\vert 0 \right\rangle _{N}=\left\vert \downarrow \downarrow \downarrow \ldots \downarrow \right\rangle _{N}$, considering the same parameters of Fig.~\ref{fig:loctermico3}(e) as a function of the number the spins. Although, $\mathcal{F}_{\mathrm{steady}}$ seems to be independent of $N$, the degree of mixture increases with $N$ as revealed by the decrease of $\mathcal{P}_{\mathrm{steady}}$, indicating the need for raising the rate $\gamma / \kappa $. As stated before, our method can be used for other  Hamiltonians as well. Thus, other entangled states can be prepared, as long as the spectrum of such Hamiltonian is resolvable, and the strongly off-ressonant terms are not coupling other modes with more than one reservoir spin ( $\sqrt{N} \lambda _{i}\ll J$ in the isotropic XY Hamiltonian). Furthermore, we can use the techniques developed to determine the inverse eigenvalue \cite{eigenvalue} and inverse eigenmode \cite{eigenmode} to engineer the Hamiltonian that leads to the desired target state and then apply our scheme.

\section{Conclusions}

In conclusion, we have proposed a simple approach for preparation of many-body entangled states in the Markovian limit. Our proposal relies on engineered dissipations assisted by a set of spins (qubits/two level systems) that mediates the dissipation and pumping of the system eigenmodes. The main requirement is that spectrum of the system has to be resolved, which is possible by a convenient choice of suitable Hamiltonians describing the spin-spin interactions, and also by allowing local control of transverse fields. Remarkably, our protocol requires neither initial state preparation nor unitary dynamics or feedback control. Moreover, the method is robust against fluctuations in the parameters as well as damping and dephasing. Within this approach, any eigenmode can be chosen as a target state. The limiting factors for scalability are the spectral resolution, which lead to individual artificial Liouvillians and the switching on/off rate $r_{g(e)}$ associated with the effective decay rates $\Gamma _{-(+)}$. Our results also suggest that the purity decrease as the number of spin in the system increases to a fixed $\gamma /\kappa$ rate. Therefore, to obtain a highly pure steady state we must increase $r_{g(e)}$, which may be experimentally  challenging. Further effort is still needed to extend our approach to embody preparation of steady states in degenerated systems as well as gapped systems in the thermodynamic limit, leading to a plethora of multipartite entangled states. Another aspect worth of further  investigation is how the non-Markovianity (when the condition $\lambda _{i}\tau <<1$ is not fulfilled) affects the preparation of  the many-body steady state.

\acknowledgments The authors  acknowledge financial support from FAPESP, CNPQ and CAPES, Brazilian agencies. 	
V.M. acknowledges the financial support of the project Fondecyt Postdoctorado $\#$3160700. E.V. also acknowledges financial support from the Brazilian agency FAPEMIG.

\end{document}